# Information collection for fraud detection in P2P financial market

*Hao* Wang[1,*], *Zonghu* Wang[1], *Bin* Zhang[2], and *Jun* Zhou[3]

[1]HC Research, HC Financial Service Group, China
[2]HC DevOps, HC Financial Service Group, China
[3]HC Big Data, HC Financial Service Group, China

**Abstract.** Fintech companies have been facing challenges from fraudulent behavior for a long time. Fraud rate in Chinese P2P financial market could go as high as 10%. It is crucial to collect sufficient information of the user as input to the anti-fraud process. Data collection framework for Fintech companies are different from conventional internet firms. With individual-based crawling request, we need to deal with new challenges negligible elsewhere. In this paper, we give an outline of how we collect data from the web to facilitate our anti-fraud process. We also overview the challenges and solutions to our problems. Our team at HC Financial Service Group is one of the few companies that are capable of developing full-fledged crawlers on our own.

## 1 Introduction

Fraud rate in Chinese P2P market is estimated over 10%. The high fraud rate poses a severe threat to P2P companies compared with conventional banking system where the fraud rate is as low as 2%. To counter the fraudulent behavior, most P2P companies create a credit-checking mechanism to block out malicious users. The credit-checking measure crawls user's information from different websites upon his authorization. Such information is crucial for the following step of anti-fraud detection which takes the collected user information as input.

Major P2P companies in China have their own information collection teams. However, many small-to-medium enterprises do not have the capacity to crawl data on their own. They rely on third-party companies to provide such services. Famous companies that provide commercial data collecting services include Tong Dun, Mo Jie, and Xin De etc. As the 3[rd] largest P2P company in China, HC Financial Service Group has built a full-fledged technical team specialized on information collection. The team collects all the necessary information of users from the internet and stores the information in the data storage platform.

Information collection for P2P products faces several challenges: 1. Websites frequently get updated, the information collection team needs to constantly monitor websites' online status; 2. Anti-crawling mechanism including security ActiveX control sometimes makes it very difficult for crawlers to collect data; 3. Information collection is very sensitive to failures. A failed attempt to collect user information blocks out the loan application process and makes the company loses a potential customer.

* Corresponding author: haow85@live.com

In this paper, we provide a description on how information collection works at HC Financial Service Group. We show the work flow of our information collection functionality and provides a brief overview of how we tackle the problems information collection team faces in general in the P2P market. For commercial privacy reasons, we could not give a thorough description of our technologies , we hope the overview we are about to outline in the following sections helps researchers and industrial workers better understand the information collection and anti-fraud processes in the industrial world.

## 2 Related work

Fraudulent behavior has a long and dark history since the advent of the internet. Tons of research has been invested to fight the malicious users targeting valuable online assets. Famous anti-fraud algorithms include Facebook-invented algorithm called CopyCatch [1] to detect the Lockstep behavior. Later Facebook team came up with a new invention called SynchroTrap [2] to detect synchronized attack. CMU researchers invented fBox [3] and SpokeEigen [4] algorithms to detect community fraud.

Fraud detection is one of the key business processes of Fintech product cycles. Common anti-fraud techniques include Bayesian Networks , Logistic Regression and other machine learning-based prediction methods rely heavily on feature engineering and domain knowledge. Vlasselaer et al. [5] provided a machine learning framework for graph-based financial fraud detection in general.

Crawlers have been widely adopted in internet companies. Google , Bing and Baidu all developed their own crawlers to collect data from the world wide web. Some crawling research has been focused on traversal strategy [6], others focus on adaptibility of crawlers [7] .

## 3 Anti-fraud data input

Anti-fraud is crucial for P2P companies and it is one of the most valuable technical assets. Anti-fraud algorithms usually take information related to users and apply feature engineering to the information before the fraud detection mechanism. The user information gathered for anti-fraud processes include the following categories.

1. Financial information : Features in this group include user's personal income , car rent , house rent , etc.
2. Work information : Features in this group include user's company's income , how long the company has been founded, etc.
3. Transaction information : Features in this group include the amount of money user borrows in the transaction , whether the user has submitted applications before , etc.
4. Demographic information: Features in this group include the number of family members of the user, etc.

There are two different sources of the information: user input and data crawling. HC Financial Service Group have both online systems and offline outlets serving customers who want to loan money from borrowers. In our online system, users open up the app and input required information in the online forms and authorize the system to crawl data from different websites. At our offline outlets , company sales staff help the customers to finish the process.

We collect user information from websites such as People's Bank of China，China Mobile etc. For commercial privacy reasons , we could not disclose the full list of the websites where we collect information from the user. Instead , we give the following list of categories of websites where we collect the user information :

1. Communication Overview : We collect user's communication information necessary for our identity authorization and credit analysis process from websites such as China Mobile.

2. Credit Report : We collect user's credit report from government agencies such as People's Bank of China.

3. Bank Account Information : We also collect user's bank account information for us to better understand the financial status of the user.

The above information is well needed in further processing of the data before the anti-fraud algorithms get involved. In the following section, we provide the workflow architecture of the information collection system.

## 4 Data collection workflow

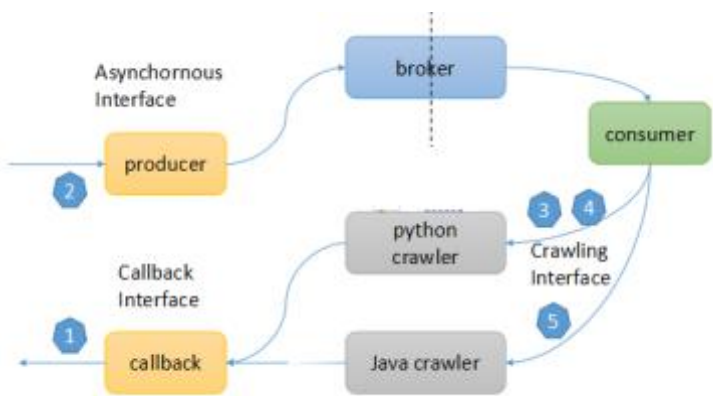

**Fig. 1.** Data collection workflow at HC financial service group.

Fig 1. illustrates the data collection workflow at HC Financial Service. The user's data crawling request is first fed into the producer. Data then flows into consumer via brokers. Then the data crawling interface chooses to use Java-written or Python-written programs to collect data from the web and return the results to the user interface.

Our crawling procedures are all individual-based, meaning that we do not do massive data collecting similar to what search engines do. Instead, we only invoke the crawling process for each user when requested. We use the most commonly used crawling tookits such as Scrapy and PhantomJS for our purposes. Although authorized crawling doesn't sound technologically sophisticated , it poses many technical challenges to our team. We give an outline of the problems we encounter in the next section.

## 5 Challenges

Since crawling data is an inalienable part of the entire P2P product cycle, it is crucial to maintain its robustness and efficiency. Although at HC Financial Service Group, we have a team of  more than 20 people working on data collection. We are still facing severe challenges that we need to deal almost routinely on a daily basis to guarantee the proper functioning of our products. To be more specific, we encounter the following problems when crawling data from the web :

1. Major Chinese banks installed ActiveX controller on the website as security guard. Commonly used crawling toolkits such as Scrapy and PhantomJS could not pass through it.

2. Complex Javascript-written encryption scripts that prevents simple crawling techniques from gathering data.

3. Crawling process intermittently gets blocked by the website.

4. Frequent rewriting and updating of the website.

5. There are too many websites from which data needs to be gathered. Each website needs code inspection and parsing manually, which consumes lots of people and time.

6. Some of our crawling toolkits have technical restrictions. For instance, PhantomJS consumption of memory is high, and it grows fast as the number of threads increase.

## 6 Solutions

ActiveX controller has been one of the biggest challenges we are facing in the data collecting team . American bank customers are not familiar with ActiveX controller for it is nearly never used in banks' websites. On the contrary, most of major banks in China utilizes ActiveX controller (Fig 2) as a necessary safeguard measure against hackers and malicious users.

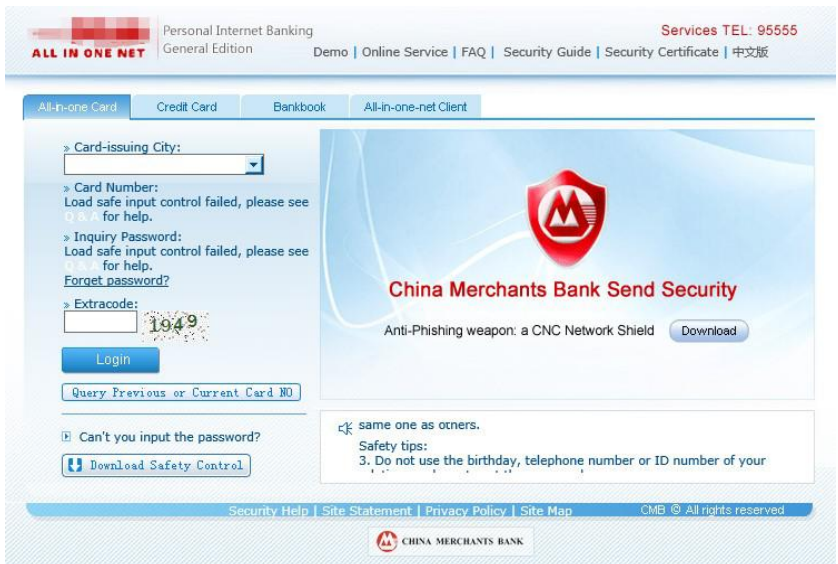

**Fig. 2.** Notification from China Merchant's Bank that ActiveX controller needs to be installed to enable login.

In our crawling system, we use success rate to measure the robustness of the system. A 90% robustness metric means among 100 requests to crawl the data, 10% of the requests failed. Websites with ActiveX controller enabled lowers down the success rate of crawling system by half compared with data collection results on other websites.

To solve this problem , we create a Windows ActiveX controller service that emulates real user's online behavior to pass through the ActiveX controller security guard. By doing so , we are able to achieve a success rate that beats most of the third-party crawling solution providers.

It is a tedious job to monitor and maintain the status of different websites. We established a monitoring system where each crawling failure would send a message notifying the developer. Whenever a website gets updated or monitored , it will trigger a failure for the crawlers and the developers will be notified instantly.

## 7 Conclusion

Fraud detection relies heavily on what users input and what we could know about the user. The more we learn about the user, the more capable we will become to predict user's credibility.Data collection is an inalienable part of the entire fraud detection process. The robustness and efficiency of crawlers is critical for the entire P2P product cycle. In this paper, we give an overview of the data collection system at HC Financial Service Group. We also outlined challenges we are facing when collecting data and how we managed to solve these issues. Although data crawling is unavoidable for P2P companies, many small-to-medium sized firms are not capable of developing data collecting systems on their own. We hope this paper could help them as well as other teams in bigger corporations to develop better systems.

Data collection might seem trivial to many experienced industry workers. This is partially due to the fact that the most mentioned crawlers are those used in search engine companies. A technical failure of the system does not lead to direct loss of money. Many crucial technical issues are hidden behind the simple and trivial appearance of the technology. However, for a financial company like HC Financial Service Group, robustness and efficiency become topmost important tasks for the crawling team. Trivial problems elsewhere pose great threats in the financial world.

In future work, we would like to keep improving our system . In particular, we would like to build more robust and efficient systems by solving the challenges mentioned in this paper. We wish our crawling system could serve as a model example of how Fintech companies should build their data collecting systems.